\newcommand{\xx}{{\bm x}}
\newcommand{\kk}{{\bm k}}
\newcommand{\pp}{{\bm p}}

\newcommand{\rr}{{\bm r}}

\newcommand{\uu}{{\bm u}}

\newcommand{\xc}{\mathcal{x}}
\newcommand{\xcb}{{\bm \xc}}
\newcommand{\kc}{\mathcal{k}}
\newcommand{\kcb}{{\bm \kc}}
\newcommand{\pc}{\mathcal{p}}
\newcommand{\pcb}{{\bm \pc}}

\newcommand{\dd}{\mathrm{d}}
\newcommand{\ee}{\mathrm{e}}

\newcommand{\PP}{\mathbb{P}}
\newcommand{\QQ}{\mathbb{Q}}

\newcommand{\WW}{{\mathcal W}}
\newcommand{\EE}{{\mathcal E}}
\newcommand{\UU}{\mathcal{U}}
\newcommand{\LL}{\mathcal{L}}

\newcommand{\joakim}[1]{\textcolor{black}{#1}}

\documentclass{jfm}
\usepackage{latexsym}
\usepackage{graphicx}
\usepackage{amssymb,amsmath,amsbsy}
\usepackage{bm}
\usepackage{epstopdf, epsfig}
\usepackage{comment}

\usepackage{dutchcal}

\usepackage{color}

\newcommand{\revision}{\color{black}}

\shorttitle{Hydrodynamic instabilities in a sheet of microswimmers}
\shortauthor{V.~\v{S}kult\'ety, D.~B\'ardfalvy, J.~Stenhammar, C. Nardini and A.~Morozov}

\title{Hydrodynamic instabilities in a 2-D sheet of microswimmers embedded in a 3-D fluid}

\author{Viktor \v{S}kult\'ety\aff{1},
	D\'ora B\'ardfalvy\aff{2},
	Joakim Stenhammar\aff{2}, \\ Cesare Nardini\aff{3}
	\and Alexander Morozov\aff{1}\corresp{\email{alexander.morozov@ed.ac.uk}}}

\affiliation{
    	\aff{1} SUPA, School of Physics and Astronomy, The University of Edinburgh, James Clerk Maxwell Building, Peter Guthrie Tait Road, Edinburgh, EH9 3FD, United Kingdom
	\aff{2}Division of Physical Chemistry, Lund University, Box 124, S-221 00 Lund, Sweden
	\aff{3}Servicede Physique de l’\'Etat Condens\'e, CNRS UMR 3680, CEA-Saclay, 91191 Gif-sur-Yvette, France}

\begin{document}
	
	\maketitle
	
	
	\begin{abstract}
		A collection of microswimmers immersed in an incompressible fluid is characterised by strong interactions due to the long-range nature of the hydrodynamic fields generated by individual organisms. As a result, suspensions of rear-actuated `pusher' swimmers such as bacteria exhibit a collective motion state often referred to as `bacterial turbulence', characterised by large-scale chaotic flows. The onset of collective motion in pusher suspensions is classically understood within the framework of mean-field kinetic theories for dipolar swimmers. In bulk 2-D and 3-D, the theory predicts that the instability leading to bacterial turbulence is due to mutual swimmer reorientation and sets in at the largest length scale available to the suspension. Here, we construct a similar kinetic theory for the case of a dipolar microswimmer suspension {\revision restricted} to a two-dimensional plane embedded in a three-dimensional incompressible fluid. This setting qualitatively mimics the effect of swimming close to a two-dimensional interface. We show that the in-plane flow fields are effectively \emph{compressible} in spite of the incompressibility of the 3-D bulk fluid, and that microswimmers on average act as sources (pushers) or sinks (pullers).
 We analyse stability of the homogeneous and isotropic state, and find two types of instability that are qualitatively different from the bulk, three-dimensional case: First, we show that the analogue of the orientational pusher instability leading to bacterial turbulence in bulk systems instead occurs at the smallest length-scale available to the system. Second, an instability associated with density variations arises in puller suspensions as a generic consequence of the effective in-plane compressibility. Given these qualitative differences with respect to the standard bulk setting we conclude that confinement can have a crucial role in determining the collective behaviour of microswimmer suspensions.
	\end{abstract}
	
	
	\begin{keywords}
		Authors should not enter keywords on the manuscript, as these must be chosen by the author during the online submission process and will then be added during the typesetting process (see http://journals.cambridge.org/data/\linebreak[3]relatedlink/jfm-\linebreak[3]keywords.pdf for the full list)
	\end{keywords}

	
	\section{Introduction}\label{sec:introduction}
	
	To move through viscous fluids, motile micro-organisms exert forces and torques on their environment. Due to their small size, the resulting motion is dominated by the viscous stresses in the fluid, and is well-described by the Stokes equation \citep{Lauga2009}. The velocity fields created by micro-organisms in this regime can thus be described by a combination of the singular solutions of the Stokes equation, which decay algebraically in space \citep{Spagnolie2012}. The leading singularity for force- and torque-free microswimmers suspended in an infinite viscous fluid is given by a force dipole, which exhibits slow spatial decay, $r^{-d+1}$, where $r$ is the distance from the microorganism and $d$ is the dimensionality of space. Such velocity fields are long-ranged, and even rather dilute suspensions of microorganisms can therefore exhibit significant fluid motion. Such `self-stirring' in the absence of an external forcing is a strongly non-equilibrium phenomenon characteristic of active matter \citep{Marchetti2013} and results in a unique set of transport and mechanical properties of microswimmer suspensions, such as enhanced diffusivity of tracer particles~\citep{Wu2000,Leptos2009,Mino2013,Jepson2013} and significant changes of the apparent suspension viscosity~\citep{Rafai2010,Lopez2015,Saintillan2018,Martinez2020}.

    Depending on the symmetry of their hydrodynamic flow, microswimmers can be divided into two distinct classes: $(i)$ `pushers', which accurately describe most swimming bacteria such as \emph{E. coli} \citep{Drescher2011}, and $(ii)$ `pullers', whose reversed hydrodynamic flow field is typically exemplified by the alga \emph{C. reinhardtii} \citep{Jeffrey2010}. While these two flow fields lead to equivalent statistical properties for very dilute microswimmer suspensions, where swimmers can be treated as effectively noninteracting, they lead to strikingly different forms of swimmer-swimmer correlations beyond this limit. Arguably, the most profound effect of such swimmer-swimmer correlations is the transition to large-scale, collective motion in elongated pusher-like microswimmers such as bacteria, often referred to as \emph{bacterial turbulence} \citep{Wensink2012a,Dunkel2013}. The mechanism of the transition to bacterial turbulence is usually rationalised based on mean-field kinetic theories that take into account the presence of long-range hydrodynamic interactions between microswimmers~\citep{Saintillan2008}. For three-dimensional suspensions of pusher bacteria, these theories predict the onset of large-scale flows above a well-defined critical microswimmer volume fraction, and identifies mutual particle reorientation due to hydrodynamic interactions as the main mechanism behind the instability, while no such instability occurs in three-dimensional puller suspensions. Even far below the onset of collective motion, the long-ranged hydrodynamic interactions result in strong correlations between microswimmers, leading to an enhancement of the effective tracer diffusivity in pusher suspensions compared to puller suspensions at the same density~\citep{Stenhammar2017,Skultety2020}.

	In practice, observations of three-dimensional bulk collective flows are complicated by the unavoidable presence of boundaries in experimental setups used to study motile microorganisms. Importantly, motile organisms accumulate in close vicinity of solid \citep{Allison2008} and liquid \citep{Vladescu2014} boundaries, significantly depleting the bulk in-between. Therefore, from an experimental perspective, it is much more convenient to study the ensuing collective motion in a layer of microswimmers confined by either type of boundary \citep{Zhang2010,Chen2012,Sokolov2012,Gachelin2014}. 
	
The theoretical description of confined microswimmer suspensions is significantly more involved than their bulk fluid counterparts discussed above, as it includes several different effects of the confinement. Firstly, the spatial decay of Stokesian singularities is modified by the presence of boundaries \citep{Spagnolie2012}, with the corresponding far-field fluid velocity being dominated by flow singularities other than the force dipole which is dominant in bulk fluids {\revision\citep{Mathijssen2016,Jeanneret2019,Brotto2013}}. Secondly, the interactions between microswimmers and the boundaries might depend on the precise shape of the microswimmers, their mechanism of propulsion, surface charge and roughness, etc. The corresponding theory would need to sufficiently resolve the near-field fluid velocity generated by the microswimmers and include appropriate steric interactions~\citep{Pessot2018,Zantop2020} and the above-mentioned effects. Such a theory would be very challenging and strongly dependent on the specific system under investigation.

	\begin{figure}
		\centering
		\includegraphics[width=0.5\linewidth]{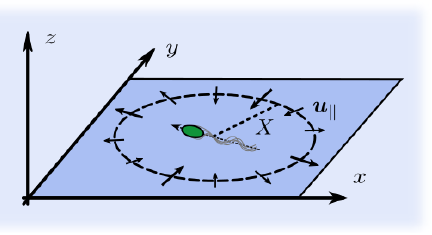}
		\caption{Schematic picture of the flow around a single pusher microswimmer {\revision restricted} to a two-dimensional plane embedded in a three-dimensional fluid. Note that the net flow $\uu_{\parallel}$ going through the dashed circle is nonzero; in
the two-dimensional plane, pushers therefore, on average, act as fluid sources, while pullers act as effective sinks.} \label{KT:fig.scheme}
	\end{figure}
	An {\revision additional} important physical effect inherent to confined systems, though rarely discussed explicitly, is the effective compressibility of the in-plane flow fields generated by a layer of microswimmers {\revision \citep{Salbreux2009,Julicher2018,Maitra2020,Huang2021,Maitra2023}. A similar effect was previously discussed in the context of hydrodynamically-interacting colloidal particles under partial confinement \citep{Bleibel2014,Bleibel2015,Bleibel2017}.} To illustrate the point, we consider a force dipole oriented parallel to a solid wall at a distance $h$ from it. The fluid velocity component $\uu_{\parallel}$ parallel to the wall can be deduced from the image system developed by \cite{Blake1971} for a point force next to a solid boundary with vanishing boundary condition $ \uu_{\parallel}(z=h) = 0 $, and is given by \cite{Spagnolie2012}:
	\begin{align}
	\uu_{\parallel} (\xcb) =&\ \frac{\kappa}{8\pi} \Bigg( \frac{\xcb}{|\xcb|^3} \left[ 3 \frac{(\xcb\cdot\pcb)^2}{|\xcb|^2} - 1 \right] + \frac{\xcb}{R^3} \nonumber
	\\
	&\hspace{2cm} - \frac{3\xcb (\xcb\cdot\pcb)^2 + 6h^2 \left\{\xcb + 2 \pcb (\xcb\cdot\pcb)\right\}}{R^5} + \frac{30 h^2 (\xcb\cdot\pcb)^2 \xcb}{R^7} \Bigg).
	\end{align}
	Here, $\pcb$ and $\xcb$ are two-dimensional vectors that lie in the plane parallel to the wall and denote respectively the dipole orientation and the point where the velocity is evaluated relative to the position of the swimmer. Furthermore, $R=\sqrt{|\xcb|^2 + 4 h^2}$, and $\kappa$ is the strength of the dipole. Next, we calculate the total flux of the fluid through a circle of radius $X$ centred on the microswimmer and parallel to the wall, as shown in Fig. \ref{KT:fig.scheme}, yielding 
	\begin{align}
	\int_{|\xcb|=X} \dd \xcb\cdot\uu_{\parallel} = \frac{\kappa}{8 X} \Biggl[ 1 - \frac{X^4-10 X^2 h^2+64h^4}{\left(X^2+4h^2 \right)^{7/2}}X^3\Biggr]. \label{Intro2d:eq.flux}
	\end{align}
	Hence, the total flux through an arbitrary circle around a microswimmer is non-zero, indicating that if we only consider the velocity components parallel to the wall, they represent an effectively compressible velocity field. Moreover, one can show that the prefactor multiplying $\kappa$ is strictly positive, and the sign of the flux is therefore determined by the sign of the dipolar strength: pushers with $\kappa>0$ on average correspond to hydrodynamic sources in the plane parallel to the wall, while pullers with $\kappa<0$ correspond to hydrodynamic sinks. Therefore, when averaged over all orientations, two pushers advect each other to maximise their mutual separation, while two pullers do the opposite. This argument illustrates that hydrodynamic interactions between dipolar microswimmers moving next to a boundary have a very different nature than their bulk counterparts, which has previously been shown to promote dynamic self-assembly and crystallisation in active particle systems \citep{Thutupalli2018,Singh2016}. We stress that this effective interaction does not correspond to extra forces or torques on the particles, which is a common assumption when describing `dry' active particles moving on a frictional substrate \citep{tenHagen2015}, as the particles immersed in a fluid are strictly force and torque free.
		
	The aim of this work is to understand the effect of this in-plane compressiblity on collective motion in a minimal setting. We consider a layer of microswimmers {\revision restricted} to move in an infinitely thin, two-dimensional plane embedded in a three-dimensional bulk fluid in the absence of any boundaries, corresponding to taking $h\to\infty$ in the example above. We assume low number density of microswimmers and approximate their velocity fields by those of three-dimensional force dipoles. As can be seen from Eq. \eqref{Intro2d:eq.flux}, even in this limit, the in-plane velocity field generated by the microswimmer is still effectively compressible. Therefore, although this specific setup is difficult to realise experimentally, it allows us to single out the effect of in-plane compressibility without needing to deal with other effects due to the confinement discussed above. 
	
	{\revision Below we demonstrate that a two-dimensional layer of microswimmers embedded in a three-dimensional fluid is unstable for both pushers and pullers. We show that pusher suspensions are prone to an orientational instability that sets in at the smallest length-scale available to the system, while puller suspensions exhibit an instability associated with density variations as a generic consequence of the effective in-plane compressibility.}
		
	
	The paper is organised as follows: In Section \ref{Sec:2}, we present a general kinetic theory describing a dilute suspension of microswimmers interacting through long-range velocity fields. In Section \ref{Sec:3} we review its predictions for the the widely studied case of microswimmers suspended in a three-dimensional bulk fluid. In Section \ref{Sec:4} we report on linear stability of a two-dimensional layer of microswimmers embedded in a three-dimensional fluid, which constitutes the main novel results of our work. We conclude by summarising our main findings in Section \ref{Sec:5}. 

	\newpage

	\section{Mean-field kinetic theory for microswimmer suspensions} \label{Sec:2}
	
	
	Mean-field kinetic theories of microswimmer suspensions have been extensively discussed in the literature \citep{Saintillan2008,Saintillan2008a,Subramanian2009,Hohenegger2010,Koch2011,Saintillan2013,Krishnamurthy2015}, and here we give only a brief summary of their general setup.
	
	A suspension of $N$ microswimmers moving autonomously through a viscous fluid is described by the one-particle distribution function $\Psi(\xx,\pp,t)$, which defines the instantaneous probability of finding a particle at a spatial position $\xx$ with an orientation given by the unit vector $\pp$. The distribution function is normalised such that
	\begin{align}
	\int \dd \xx\, \dd \pp \ \Psi(\xx,\pp,t) = 1.
	\label{KT:eq.NORM}
	\end{align}
	Its time evolution is assumed to be governed by the Smoluchowski equation
	\begin{align}
	\partial_{t} \Psi + \nabla^{\alpha} \{ \dot{x}^{\alpha} \Psi \} + \partial^{\alpha} \{ \dot{p}^{\alpha} \Psi \} = - \lambda \Psi + \lambda  \int \frac{ \dd \pp }{\Omega_d} \ \Psi, 
	\label{KT:eq.SE}
	\end{align}
	where Greek superscripts denote Cartesian components of vectors, $\nabla^{\alpha} = \partial/\partial x^{\alpha}$, $\partial^{\alpha} = (\delta^{\alpha\beta} - p^{\alpha}p^{\beta})\partial / \partial p^{\beta}$, and $\delta^{\alpha\beta}$ denotes the Kronecker delta; $\Omega_d$ is the surface area of a $d$-sphere of unit radius, where $d$ is the dimensionality of space, and $\lambda$ is the tumbling frequency as further described below. The deterministic part of the single-swimmer dynamics is  described by the following microscopic equations of motion:
	\begin{align}
	\dot{x}^{\alpha} &= v_{s} p^{\alpha} + \UU^{\alpha}(\xx),  
	\label{KT:eq.SEx} \\
	\dot{p}^{\alpha} &= (\delta^{\alpha\beta} - p^{\alpha}p^{\beta}) (\WW^{\beta\gamma}(\xx) + B \EE^{\beta\gamma}(\xx)) p^{\gamma},
	\label{KT:eq.SEp}
	\end{align}
	where dots denote time derivatives. According to Eq.\eqref{KT:eq.SEx}, each particle changes its spatial position due to its own swimming with a constant speed $v_s$ in the direction of its orientation $\pp$ and due to advection by the fluid flow with velocity $\bm{\mathcal U}$ at its position $\xx$, created by all other microswimmers. Microswimmer orientations change according to {\revision Jeffery’s} equation, \eqref{KT:eq.SEp}, which describes particle rotation by the gradients of the fluid velocity at its position \citep{kimkarrila}. Here, $ \WW^{\alpha\beta}(\xx) = \tfrac{1}{2} (\nabla^{\beta} \UU^{\alpha}(\xx) - \nabla^{\alpha} \UU^{\beta}(\xx)), $ and $ \EE^{\alpha\beta}(\xx) = \tfrac{1}{2} (\nabla^{\beta} \UU^{\alpha}(\xx) + \nabla^{\alpha} \UU^{\beta}(\xx)) $ are the vorticity and the rate-of-strain tensors. The Bretherton parameter $B$ encodes the shape of the particle \citep{kimkarrila}, with $B=1$ and $B=0$ respectively corresponding to needle-like and spherical particles. Finally, each microswimmer randomly selects a new orientation (tumbles) with rate $\lambda$. This discrete stochastic process cannot be incorporated in the time-evolution Eqs. \eqref{KT:eq.SEx} and \eqref{KT:eq.SEp} in a straightforward manner, but is readily described within the kinetic theory \citep{Subramanian2009,Koch2011}, as given on the right-hand-side of \eqref{KT:eq.SE}. 
	
	The final ingredient of the theory is provided by the relationship between the local velocity field and the one-particle distribution function, given by
	\begin{align}
	\UU^{\alpha}(\xx,t) = N \int \dd \xx' \dd \pp' \ u^{\alpha}(\xx-\xx',\pp') \Psi(\xx',\pp',t).
	\end{align}
	Here, $\uu(\xx-\xx',\pp')$ is the microscopic velocity field created at $\xx$ by a microswimmer located at $\xx'$ with the orientation $\pp'$. In the following, we approximate $\uu$
	by the dipolar field generated by two equal and opposite point forces applied to the fluid infinitesimally close to each other \citep{Lauga2009}. While this is a good approximation for dilute suspensions of microswimmers in three-dimensional bulk systems \citep{Lauga2009,Spagnolie2012,Skultety2020}, its validity in the context of the present study is discussed in Section \ref{Sec:5}.

	\subsection{Linear stability of the homogeneous and isotropic state}
	
	The Smoluchowski equation~\eqref{KT:eq.SE} and the normalisation condition \eqref{KT:eq.NORM} admit as a solution the homogeneous and isotropic state $\Psi_{\text{HI}} = 1/( \Omega_d V_d )$, where $V_d$ is the $d$-dimensional volume of the suspension. The existence of this solution relies on the condition
	\begin{align}
	\int \dd \xx' \dd \pp' \ u^{\alpha}(\xx-\xx',\pp') = 0,
	\end{align}
	which is true in all cases considered in this work. Notice that, while integrals over $ \xx' $ and $ \pp' $ vanish independently in two- and three-dimensional bulk suspensions, only the positional integral vanishes for microswimmers {\revision restricted} to a two-dimensional plane in a three-dimensional fluid. 
	
	Stability of the homogeneous and isotropic state is determined by time evolution of small perturbations around $\Psi_{\text{HI}}$. Techniques to study such problems are well-established and have been extensively applied to linear stability analysis of the homogeneous and isotropic state for dilute, three-dimensional suspensions of microswimmers \citep{Saintillan2008, Saintillan2008a,Subramanian2009,Hohenegger2010}. Here, we use a somewhat different methodology \citep{Stenhammar2017,Martinez2020}, which is better adapted for our purpose.
	
	First, we introduce a perturbation $\delta \Psi(\xx,\pp,t)$ of the one-particle distribution function around the homogeneous and isotropic state, \emph{i.e.},
	\begin{align}
	\Psi(\xx,\pp,t) = \frac{1}{\Omega_d V_d} + \delta \Psi(\xx,\pp,t),
	\end{align}
	where we assume that $\delta \Psi(\xx,\pp,t)$ is small and that its integral over $\xx$ and $\pp$ vanishes. In what follows, we employ the $d$-dimensional Fourier transform defined through
	\begin{align}
	\hat{f}(\kk)  &= \int \dd \xx \ f(\xx) \ee^{-i\kk\cdot\xx}, \quad f(\xx) = \frac{1}{(2\pi)^{d}} \int \dd \kk \ \hat{f}(\kk) \ee^{i\kk\cdot\xx}, 
	\label{KT:eq.FT}
	\end{align}
	where $f$ is an arbitrary function of $\xx$, and $\hat f$ denotes its Fourier space representation. The linearised Smoluchowski Eq.\eqref{KT:eq.SE} in Fourier space reads
	\begin{align}
	\Bigl[ \partial_{t}  + \lambda + i v_{s} \pp\cdot\kk \Bigr] \delta \hat{\Psi} &= \frac{\lambda}{\Omega_d}  \delta \hat{\rho}  +  \frac{ n}{\Omega_d} \Bigl[ d B p^{\alpha} p^{\beta}  - (1+B) \delta^{\alpha\beta}  \Bigr] ik^{\alpha} \delta \hat{\UU}^{\beta}, 
	\label{LS:eq.SELin}
	\end{align}
	where $n=N/V_d$ is the number density of particles, and we have defined the perturbations of the microswimmer density and fluid velocity as
	\begin{align}
	\delta \hat{\rho}(\kk,t) &= \int \dd \pp \ \delta \hat{\Psi}(\kk,\pp,t), \\
	\delta \hat{\UU}^{\alpha}(\kk,t) &= \int \dd \pp \ \hat{u}^{\alpha}(\kk,\pp) \delta \hat{\Psi}(\kk,\pp,t).
	\end{align}
	We proceed by assuming an exponential solution to \eqref{LS:eq.SELin} of the following form
	\begin{align}
	\delta \hat{\Psi}(\kk,\pp,t) = \delta \hat{\Psi}(\kk,\pp) \ee^{\chi t},  
	\label{LS:eq.SELinExp}
	\end{align}
	where the sign of the real part of the temporal eigenvalue $\chi$ determines stability of the system against infinitesimal perturbations: For Re$[\chi]<0$ the system is linearly stable, while for Re$[\chi]>0$ the system is linearly unstable. It should be noted that a more general approach to the temporal linear stability analysis is to treat the problem as an initial value one and obtain the solution via the Laplace transformation, as was done in \citep{Stenhammar2017,Skultety2020}. Working with the Ansatz \eqref{LS:eq.SELinExp} is mathematically simpler but may not necessarily produce a dispersion law for all parameter values, as we discuss further below.

	
	Inserting \eqref{LS:eq.SELinExp} into \eqref{LS:eq.SELin}, we derive the following closed set of equations for the density and velocity perturbations
	\begin{align}
	\delta \hat{\rho} &= i n \delta \hat{\UU}^{\alpha}  \int  \frac{\dd \pp}{\Omega_d} \  \frac{ B d p^{\alpha} \pp\cdot\kk - (1+B)k^{\alpha} }{\LL(\chi,\pp\cdot\kk)} + \delta \hat{\rho} \int \frac{\dd \pp}{\Omega_d} \ \frac{\lambda}{\LL(\chi,\pp\cdot\kk)}, \label{LS:eq.LinEq1} \\
	\delta \hat{\UU}^{\alpha} &= i n \delta \hat{\UU}^{\beta}  \int \frac{\dd \pp}{\Omega_d} \  \frac{ \hat{u}^{\alpha}(\kk,\pp) [B d p^{\beta} \pp\cdot\kk - (1+B)k^{\beta}] }{\LL(\chi,\pp\cdot\kk)} + \delta \hat{\rho} \int \frac{\dd \pp}{\Omega_d} \ \frac{\lambda \hat{u}^{\alpha}(\kk,\pp)}{ \LL(\chi,\pp\cdot\kk) }, 
	\label{LS:eq.LinEq2} 
	\end{align}
	where $\LL(\chi,\pp\cdot\kk) = \chi + \lambda + i v_s \pp\cdot\kk$. The solution to this eigenvalue problem depends on the precise form of the microscopic velocity field $\hat{\uu}(\kk,\pp)$, which, in turn, depends on the dimensionality of space and the dimensionality of the vectors $\pp$ and $\kk$. In the following, we begin by revisiting the known results on linear stability of bulk three-dimensional suspensions (\cite{Saintillan2008,Saintillan2008a}) before turning our attention to a two-dimensional layer of microswimmers embedded in a three-dimensional fluid.
	
		\section{Microswimmers in an unbounded bulk suspension} 
	\label{Sec:3}
	
	\noindent In this Section, we consider the case of microswimmers that move freely in a 3-D space filled with an incompressible fluid. The dipolar velocity field for $d=3$ is given by \citep{Lauga2009} 
	\begin{align}
	u_{3d}^{\alpha}(\xx,\pp) = \frac{\kappa}{8\pi} \frac{x^{\alpha}}{|\xx|^{3}} \left[ 3 \frac{(\xx\cdot\pp)^{2}}{|\xx|^{2}} - 1 \right], 
	\label{LS3d:eq.u3d}
	\end{align}
	and, as shown in Appendix \ref{App:ud}, its Fourier transform is
	\begin{align}
	\hat{u}_{3d}^{\alpha}(\kk,\pp) = - i \kappa \frac{\kk\cdot\pp}{k^{2}} \PP^{\alpha\beta} p^{\beta}. 
	\label{LS3d:eq.u3dK}
	\end{align}
	Here, $\kappa$ is the dipolar strength and $ \PP^{\alpha\beta} = \delta^{\alpha\beta} - k^{\alpha} k^{\beta}/k^{2} $ is the transversal projection operator that ensures that Eq.\eqref{LS3d:eq.u3dK} satisfies the incompressibility condition $k^\alpha \hat{u}_{3d}^\alpha = 0$. Setting $d = 3$ in \eqref{LS:eq.LinEq1}--\eqref{LS:eq.LinEq2}, using the fact that $\delta \hat{\UU}^{\alpha} = \PP^{\alpha\beta} \delta \hat{\UU}^{\beta}$, and assuming that $ \chi $ does not depend on the orientation $\pp$ (see further discussion below), we obtain the following set of equations \citep{Martinez2020}
	\begin{align}
	\delta \hat{\rho} &= \frac{\lambda}{v_{s} k} \arctan(b) \delta \hat{\rho}, 
	\label{LS3d:eq.rho} \\
	\delta \hat{\UU}^{\alpha} &= \frac{B n \kappa}{v_{s}k} \frac{b(3+2b^{2}) - 3 (b^{2} + 1) \arctan(b)}{2b^{4}} \delta \hat{\UU}^{\alpha}, 
	\label{LS3d:eq.U}
	\end{align}
	where we have introduced the dimensionless parameter
	\begin{align}
	b = \frac{v_{s}k}{\chi + \lambda}. 
	\label{LS3d:eq.b}
	\end{align}
	We note that Eqs. \eqref{LS3d:eq.rho} and \eqref{LS3d:eq.U} are decoupled, and can be studied independently. The former equation involves only $\delta \hat{\rho}$, and we thus refer to it as the density eigenvalue problem. The latter equation is known to lead to a long-wavelength instability associated with orientational degrees of freedom only \citep{Saintillan2008,Saintillan2008a,Hohenegger2010,Subramanian2009,Stenhammar2017}, and we thus refer to it as the orientational eigenvalue problem. We will now study these two problems separately, while the analogous calculation for the case of an infinite 2-D suspension is carried out in Appendix \ref{App:2D} with qualitatively identical conclusions.
	
	\subsection{Bulk orientational instability}  \label{LS2D:eq.OrientInst}
	
	\begin{figure}
		\centering
		\includegraphics[width=0.9\linewidth]{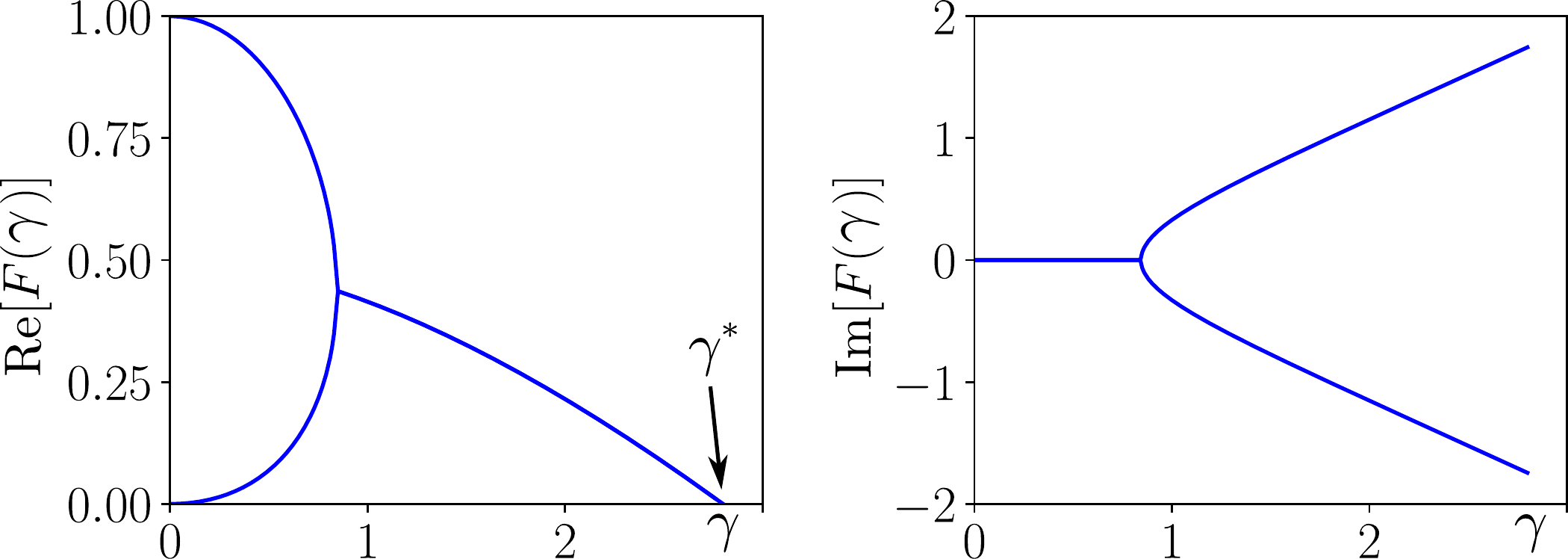}
		\caption{The real (left) and imaginary (right) part of the function $ F(\gamma) $ from \eqref{LS3d:eq.sol}. } 
		\label{LS3d:fig.1}
	\end{figure}
	
	Eq.\eqref{LS3d:eq.U} yields the following eigenvalue problem:
	\begin{align}
	\gamma = \frac{5}{2} \frac{b(3+2b^{2}) - 3(b^{2} + 1) \arctan(b)}{b^{4}}, \quad \gamma \equiv \frac{5v_{s}k}{Bn\kappa}.
	\label{LS3d:eq.aF}
	\end{align}
    We now solve Eq. \eqref{LS3d:eq.b} for $\chi$ and use the definition of $\gamma$ from Eq. \eqref{LS3d:eq.aF}, to arrive at
 	\begin{align}
	\chi = - \lambda + \frac{1}{5} F(\gamma) B \kappa n, \quad F(\gamma) = \frac{\gamma}{b(\gamma)} .
	\label{LS3d:eq.sol}
	\end{align} 
	Eq. \eqref{LS3d:eq.aF} can be inverted numerically to obtain $b(\gamma)$ or, equivalently, $F(\gamma)$, while an analytical approximation to $F(\gamma)$ was developed in \cite{Skultety2020} and \cite{Martinez2020}. In Fig. \ref{LS3d:fig.1} we plot the results of our numerical evaluation of $F(\gamma)$, showing that $(i)$ the real part of $F$ is positive and decreases monotonically with increasing $\gamma$, and $(ii)$ no solution is found for $\gamma>\gamma^*$, a fact that is discussed further below. Together with Eq.~\eqref{LS3d:eq.sol}, this implies that puller suspensions ($\kappa < 0$) are always stable while the instability of pusher suspensions ($\kappa > 0$) sets in at the largest possible length-scale, corresponding to $\gamma \rightarrow 0$ and $F(\gamma) \rightarrow 1$. It follows from Eq.~\eqref{LS3d:eq.sol}, and from imposing $\chi>0$, that pusher suspensions are unstable for densities larger than 
	\begin{equation}
	    n_{c} = \frac{5\lambda}{B\kappa}. \label{LS3d:eq.nc}
	\end{equation}
    This latter result is exact, and can be alternatively derived directly from \eqref{LS3d:eq.U} by setting $k \rightarrow 0$.
    
The origin of this instability is the mutual reorientation of microswimmers~\citep{Saintillan2008,Saintillan2008a,Hohenegger2010,Subramanian2009,Stenhammar2017}. It is thus unsurprising to observe that no instability is present for spherical particles ($B=0$). Furthermore, the instability condition \eqref{LS3d:eq.nc} does not involve the swimming speed $v_s$, implying that the orientational instability persists even for \emph{shakers} -- microswimmers that do not self-propel, yet are capable of generating dipolar fields~\citep{Stenhammar2017}. 
	
	
	We finally comment on the fact that no solution for $F(\gamma)$ is found for $\gamma>\gamma^*\approx 2.8$. This means that a solution to the linearised kinetic equation \eqref{LS:eq.SELin} obeying the Ansatz \eqref{LS:eq.SELinExp} ceases to exist for $\gamma>\gamma^*$. The reason for this is that, in the process of deriving \eqref{LS3d:eq.rho}--\eqref{LS3d:eq.U}, we have assumed that $\chi$ is independent of the swimmer orientation $\pp$. \cite{Hohenegger2010} performed a detailed analysis of the eigenvalue problem in this parameter range avoiding such an assumption, and concluded that the system is linearly stable for $\gamma > \gamma^{*}$. The most unstable eigenvalue is thus given by \eqref{LS3d:eq.sol}, leading to the instability threshold \eqref{LS3d:eq.nc}. The same conclusion is reached by solving \eqref{LS:eq.SELin} using Laplace transform techniques as was done in \cite{Stenhammar2017,Skultety2020}.
	
	\subsection{Density fluctuations in bulk suspensions} \label{LS2D:eq.DebInst}
	
	\noindent Both puller and pusher suspensions are stable against density fluctuations. For wavenumbers $k \leq \pi\lambda/(2v_{s}) $ this can be readily shown from Eq.~\eqref{LS3d:eq.rho}, yielding 
 	\begin{align}
	\chi = - \lambda + \frac{v_{s}k}{\tan(v_{s}k/\lambda)}.
	\label{LS3d:eq.rho2}
	\end{align}
    from which we conclude that $\chi<0$. The corresponding eigenstates represent spatial modulation of the microswimmer density, while their orientations remain isotropically distributed regardless of their spatial positions.
   Similar to the orientational instability, Eq.~\eqref{LS3d:eq.rho} does not allow one to determine stability of the suspension when $k > \pi\lambda/(2v_{s})$, since in this regime Eq.~\eqref{LS3d:eq.rho} has no solution. As discussed in Section~\ref{LS2D:eq.OrientInst}, this is a limitation of the Ansatz \eqref{LS:eq.SELinExp}. In Appendix \ref{App:2D.LT}, we analyse this problem in some detail by solving Eq.~\eqref{LS:eq.SELin} using Laplace transform techniques and show that the system is indeed stable with respect to infinitesimal density perturbations for all $k$.
	
	\section{Microswimmers {\revision restricted} to a 2-D plane} 
	\label{Sec:4}
	
	We now turn our attention to the main problem of this study -- the case of a 2-D layer of microswimmers embedded in a 3-D bulk fluid. As discussed in Section~\ref{sec:introduction}, this spatial arrangement of microswimmers leads to the in-plane fluid velocity field being effectively compressible, with pushers acting on average as sources, while pullers act as sinks. Here, we study the consequences of this effective compressibility on the onset and type of collective motion expected in this arrangement of microswimmers.
	
	The Fourier representation of the in-plane velocity field created by a microswimmer is given by (see Appendix \ref{App:ud} for details)
	\begin{align}
	u_{\text{plane}}^{\alpha}(\kcb,\pcb) = - \frac{i\kappa}{2} \frac{\kcb\cdot\pcb}{\kc} \left[ \PP^{\alpha\beta} + \frac{1}{2} \QQ^{\alpha\beta} \right] \pc^{\beta}, 
	\label{LS2d:eq.u2dK}
	\end{align}
	where $\kappa$ is the dipolar strength which has the same dimensions as in the bulk 3-D case, and $  \kcb = \left( \kc_x, \kc_y\right)$, $\pcb = \left( \pc_x, \pc_y\right)$ are respectively the in-plane wave and orientation vectors. In writing Eq.~\eqref{LS2d:eq.u2dK}, we explicitly separated the term proportional to the longitudinal projection operator $ \QQ^{\alpha\beta} = \kc^{\alpha} \kc^{\beta} / \kc^{2} $ to stress the compressible nature of $\uu_{\text{plane}}$. We now define the transversal (incompressible) and longitudinal (compressible) velocity perturbations by
	\begin{align}
	\delta \hat{\UU}_{\perp}^{\alpha} = \PP^{\alpha \beta} \delta \hat{\UU}^{\beta}, \quad \delta \hat{\UU}_{\parallel}^{\alpha} = \QQ^{\alpha \beta} \delta \hat{\UU}^{\beta},
	\end{align}
	where $ \delta \hat{\UU}^{\alpha} = \delta \hat{\UU}_{\perp}^{\alpha} + \delta \hat{\UU}_{\parallel}^{\alpha} $. Substituting \eqref{LS2d:eq.u2dK} for $\uu_{\text{plane}}$ into \eqref{LS:eq.LinEq1}--\eqref{LS:eq.LinEq2} results in the following compact set of equations determining linear stability of the system:
	\begin{align}
	\begin{pmatrix}
	M_{11}-1 & 0 & 0 \\
	0 & M_{22}-1 & i \kc^{\alpha} M_{23} \\
	0 &  i \kc^{\alpha}  M_{32} & M_{33}-1 
	\end{pmatrix}
	\cdot
	\begin{pmatrix}
	\delta \hat{\UU}_{\perp}^{\alpha} \\
	\delta \hat{\UU}_{\parallel}^{\alpha} \\
	\delta \hat{\rho}
	\end{pmatrix}
	= 0, \label{LS:eq.M}
	\end{align}
	where 	
		\begin{align}
		M_{11} &= \frac{\kappa n}{v_{s}} \frac{B \left(b^2-2 \sqrt{b^2+1}+2\right) }{2 b^3}, \quad M_{22} = \frac{\kappa n}{v_{s}} \frac{B b^2 + (b^{2} + 2B)\left( 1 - \sqrt{b^{2} + 1} \right)}{4 b^3 \sqrt{b^2+1}}, \label{LS:eq.M2} \\
		M_{23} &= \frac{\kappa\lambda}{ v_{s} \kc^{2} } \frac{1}{4 b} \left( \frac{1}{\sqrt{b^2+1}}-1 \right)  , \quad M_{32} = \frac{n}{ v_{s} \kc} \frac{2 B \left(\sqrt{b^2+1}-1\right) -b^2 (B+1)}{b \sqrt{b^2+1}} , \nonumber \\
		M_{33} &= \frac{\lambda}{\kc v_{s}} \frac{b}{\sqrt{1+b^{2}}}, \nonumber
		\end{align}
      and 
      \begin{align}
        b = \frac{v_s \kc}{\chi + \lambda}.
      \end{align}
	 As is apparent from Eq. \eqref{LS:eq.M}, $\delta \hat{\UU}_{\perp}^{\alpha}$ is decoupled from $\delta \hat{\UU}_{\parallel}^{\alpha}$ and $ \delta\hat{\rho} $. This allows us to analyse the two subsets of equations independently; as we show below, they correspond to different types of instabilities.

		\subsection{Orientational instability} \label{sec:orientational_instability}
    
    We start by considering the transversal mode $ \delta \UU_{\perp}^{\alpha} $, which amounts to solving $ M_{11} = 1 $. 
    This yields
		\begin{align}
		\frac{b^2-2 \sqrt{b^2+1}+2 }{b^3} = \frac{2v_{s}}{B \kappa n}.
		\label{LS2d:eq.ab}
		\end{align}
  Assuming that $b \neq 0$, Eq. \eqref{LS2d:eq.ab} can be transformed into a polynomial equation that can be solved analytically. Inserting the obtained roots into Eq.~\eqref{LS2d:eq.ab} we found that only two solutions to Eq. \eqref{LS2d:eq.ab} exist, and only within the region $2\sqrt{2}v_{s} \le B \kappa n $. In analogy with Eq.~\eqref{LS3d:eq.aF} for the bulk suspension, the eigenvalue $\chi$ within this parameter range is found to have the following form:
		\begin{align}
		\chi = - \lambda + G\left(\frac{2 v_{s}}{B\kappa n}\right) \frac{B \kappa n \kc}{2}, 
		\label{LS2d:eq.yd} 
		\end{align}
		where 
    	\begin{align}
    	G(x) &= 6x^{2} \left( 4 - \frac{\left(1\pm i \sqrt{3}\right)}{ H(x)} - \left(1\mp i \sqrt{3}\right) H(x) \right)^{-1}, 
    	\label{LS2d:eq.G}\\
    	H(x) &= \left[ 54 x^2 -1 + 6 \sqrt{3}x \sqrt{27 x^2 - 1}\right]^{1/3}.
    	\end{align}
Similar to the bulk suspension, in the region $ 2\sqrt{2} v_{s} > B \kappa n $, where the solution to \eqref{LS2d:eq.ab} does not exist, stability analysis requires the use of an alternative approach. This is discussed in Appendix \ref{App:2D.LT_2d}, where we show that the system remains stable in this parameter range.
     

 The function $G$ has the same qualitative features as the function $F$ in the case of bulk suspensions (compare Figs. \ref{LS3d:fig.1} and \ref{LS2d:fig.G} {\revision (see Appendix \ref{App:2D})}): its real part is positive and its largest branch is a monotonically decreasing function of its argument. Because Eq.~\eqref{LS2d:eq.yd} is an increasing function of $\kc$, the orientational instability in a two-dimensional layer of microswimmers embedded in a three-dimensional fluid sets in at the \emph{largest} possible value of $\kc$, in contrast with bulk pusher suspensions. In the absence of any physical restrictions on the maximum value taken by $\kc$, this instability occurs for
    \begin{align}
	n > n_{c} = 2\sqrt{2} \frac{v_{s}}{B \kappa}. \label{LS2d:eq.yd2}
	\end{align}
    Just as in bulk systems, the fact that $n_{c}$ diverges for spherical particles ($B=0$) shows that this is an instability driven by particle reorientations due to hydrodynamic interactions. 
{\revision However, unlike the 3-D bulk case~\eqref{LS3d:eq.nc}, $n_c$ is now a function of the swimming speed $v_s$ rather than of the tumbling rate $\lambda$, meaning that increasing the swimming speed will stabilise a sheet of pushers, which become unstable for any density in the shaker limit $v_s \to 0$. While the stabilising role of swimming in the stability and pre-transitional correlations of finite microswimmer systems has previously been discussed for 3-D bulk suspensions \citep{Saintillan2013,Liu2019,Skultety2020,Martinez2020,Albritton2023preprint}, its effect on the stability of a sheet of pusher microswimmers, Eq.\eqref{LS2d:eq.yd2}, is significantly stronger. Finally, the independence of $n_{c} $ of the tumbling rate implies that even a suspension of straight swimmers has a non-vanishing critical density, in contrast to the case encountered in three-dimensional bulk suspensions.}

	The above reasoning for deriving the orientational instability assumes that the most unstable wave vector is $\kc_c \to \infty$; this is clearly unphysical since it corresponds to microscopic length scales. In practice, the instability instead sets in at some finite length scale $l_c$, corresponding to a finite value of $\kc_c$. 
	As observed previously for three-dimensional bulk suspensions \citep{Saintillan2008,Subramanian2009} one possible source of regularisation is provided by the spatial diffusivity of microswimmers. 
    	Repeating the above analysis in the presence of Brownian diffusion, yields an additional term, $-D\kc^2$, on the right-hand side of \eqref{LS2d:eq.yd}, where $D$ is the diffusion constant. 
    Minimising the real part of the eigenvalue with respect to $k$ leads to the lengthscale $l_c \sim \sqrt{D/\lambda}$ being selected at the instability. Using the approximate values $D\sim 0.2-0.4$ $\mu$m$^2$/s \citep{Poon2013} and $\lambda \sim 1$ s$^{-1}$, this lengthscale becomes comparable with the microswimmer size, $l_c \sim 1$ $\mu$m, which is an unphysically small length scale for the instability to occur. It is therefore unlikely that spatial diffusivity is the relevant mechanism of the length scale selection in \eqref{LS2d:eq.yd}. 

	The second relevant microscopic length scale in the problem is the (density-dependent) average distance between  microswimmers, below which the system can no longer be viewed as a continuum. Thus, we assume that the length scale selected at the instability is now $l_c \sim 2/\sqrt{\pi n_c}$, leading to a maximum wave vector
	\begin{align}
	\kc_c =  \frac{2\pi}{l_c} = \sqrt{ \pi^{3} n_c }. 
	\label{LS2d:eq.lc}
	\end{align}
	The calculation of the corresponding instability threshold is outlined in Appendix \ref{App:2d_approx}, leading to the following approximate expression for $n_c$:
	\begin{align}
	n_{c} \approx \frac{4}{\pi} \left( \frac{\lambda}{B\kappa} \right)^{2/3} + 2\sqrt{2} \frac{v_{s}}{B\kappa}. \label{LS2d:eq.OrientPhic1}
	\end{align}
	We observe that the critical density \eqref{LS2d:eq.OrientPhic1} differs from its $\kc_c \to \infty$ counterpart Eq. \eqref{LS2d:eq.yd2} by the presence of the first term, which dominates at low swimming speeds. Using the experimental values $ v_s \approx 15$ $\mu$m/s, $\kappa \approx 800$ $\mu$m$^3$/s measured in \cite{Drescher2011}, together with $B \approx 1$ and $\lambda \approx 1$ s$^{-1}$, Eqs. \eqref{LS2d:eq.OrientPhic1} and \eqref{LS2d:eq.lc} give the critical length-scale $l_c\sim 4$ $\mu$m, which is larger than, although comparable to, the value resulting from translational diffusion. 

	\subsection{Density instability} \label{sec:densityinstability}

	\begin{figure}
		\centering
		\includegraphics[width=0.9\linewidth]{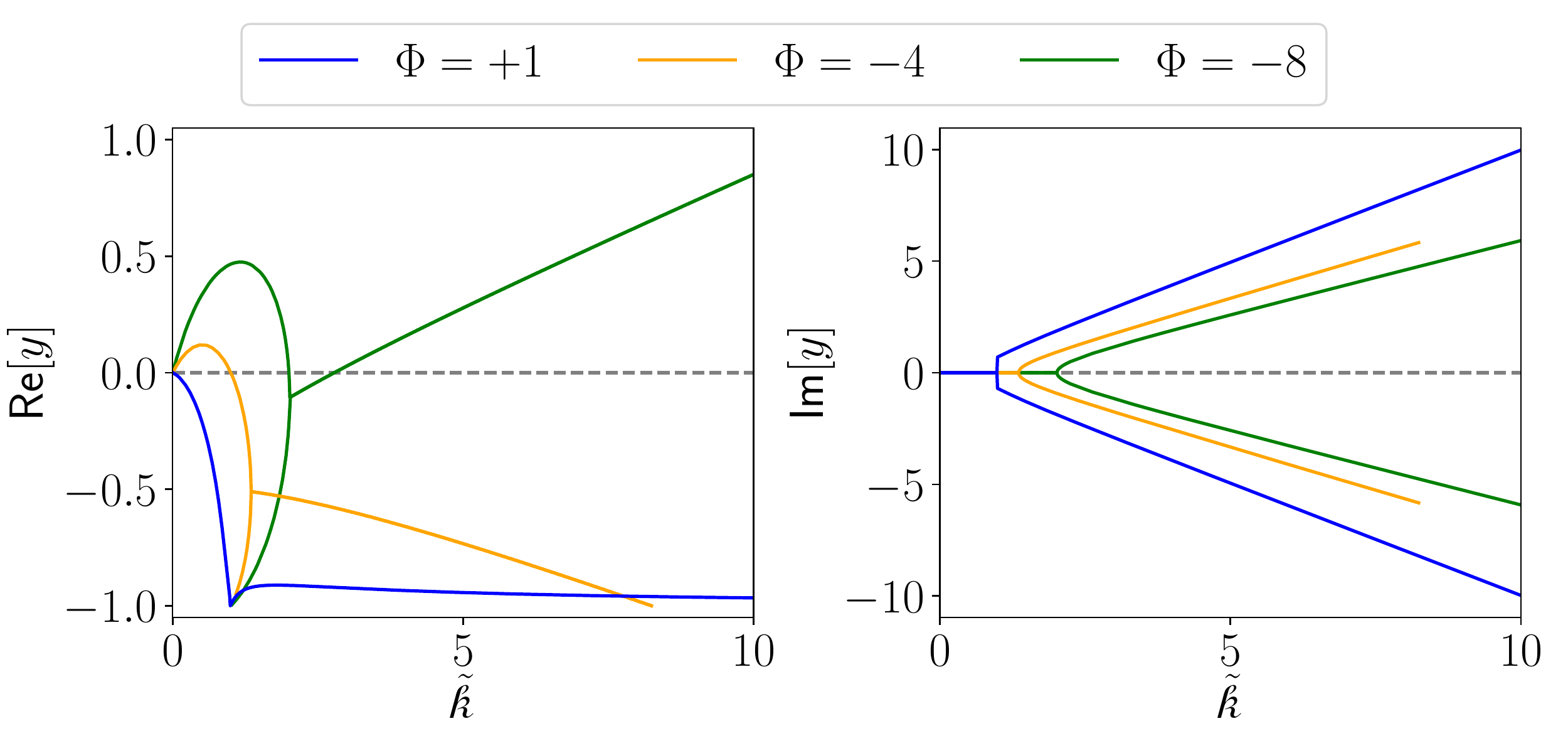}
		\caption{ The real (left) and imaginary (right) part of the eigenvalue $y = \chi/\lambda$ corresponding to density perturbations, obtained by numerically solving Eq. \eqref{LS2d:eq.denEQ}. All plots correspond to needle-like particles ($B = 1$), but are qualitatively similar for all values of $B$. For pushers ($\Phi > 0$), the real part of the eigenvalue is strictly negative. For pullers ($\Phi < 0$), the real part of the eigenvalue becomes positive at small wave vectors. At larger densities, the global maximum of $\mathrm{Re} [y]$ moves from small $\tilde{\kc}$ to $\tilde{\kc} \to \infty$. No solution exists in the region Re$ [y] < - 1 $, whose stability is instead addressed in Appendix \ref{App:2D.LT}.} \label{LS2d:fig.3}
	\end{figure} 
	\noindent The second subset of \joakim{Eqs. \eqref{LS:eq.M2}} couples $ \delta \hat{\UU}_{\parallel}^{\alpha} $ and $ \delta \hat{\rho} $ and thus governs the appearance of density modulations within the microswimmer layer. After some algebra, this set of equations can be re-expressed as 

	\begin{align}
	    \frac{\Phi \tilde{\kc} \left[ B \left(b^2 + 2- 2 \sqrt{b^2+1}\right) y -  b^2     \left(\sqrt{b^2+1}-1\right) (y+1) \right]}{4 b^4 (y+1) \left[ \sqrt{b^2+1}(y + 1)-1  \right]} = 1,
    \label{LS2d:eq.denEQ}
	\end{align}
	where we have introduced the following dimensionless quantities:
	\begin{align}
	    b &= \frac{\tilde{\kc}}{y + 1}, \quad y = \frac{\chi}{\lambda}, \quad  \Phi = \frac{\kappa}{v_{s}}n, \quad \tilde{\kc} = \frac{v_{s}}{\lambda} \kc. \label{LS2d:eq.den_DimPar}
	\end{align}
    A full analytical solution of Eq. \eqref{LS2d:eq.denEQ} cannot be found, but can be achieved perturbatively for small $\tilde{\kc}$, yielding
	\begin{align}
		y &= - \frac{1}{8} \Phi \tilde{\kc} - \left( \frac{1}{2} + \frac{B \Phi^{2}}{128} \right) \tilde{\kc}^{2} + \mathcal{O}(\tilde{\kc}^{3}). \label{LS2d:eq.den_DL}
	\end{align}
    From this expression, we conclude that $(i)$ pusher suspensions ($\Phi>0$) are stable against density modulations at large spatial scales since $y < 0$ for small $\tilde{\kc}$, and $(ii)$ puller suspensions ($\Phi < 0$) are linearly unstable at any density. Note that this instability is independent of $B$, and hence occurs even for spherical particles. Even though the puller density instability occurs at vanishingly small densities in an infinite system, for a finite system with linear size $H$ one can show that puller suspensions are unstable against large-scale perturbations only above a critical density $n_c$ given by
	\begin{align}
		n_c = -\frac{8\pi}{H} \frac{v_{s}^{2}}{\lambda\kappa}. \label{LS2d:eq.denPhic}
	\end{align}
    This result is obtained by assuming that the eigenvalue $y$ is real for small $\tilde{\kc}$, setting $y = 0$ in \eqref{LS2d:eq.denEQ}, and assuming that the critical wave vector $\kc_c$ is set by the system dimensions, \emph{i.e.}, $\kc_c = 2\pi/H$.

\begin{figure}
			\centering
			\includegraphics[width=1\linewidth]{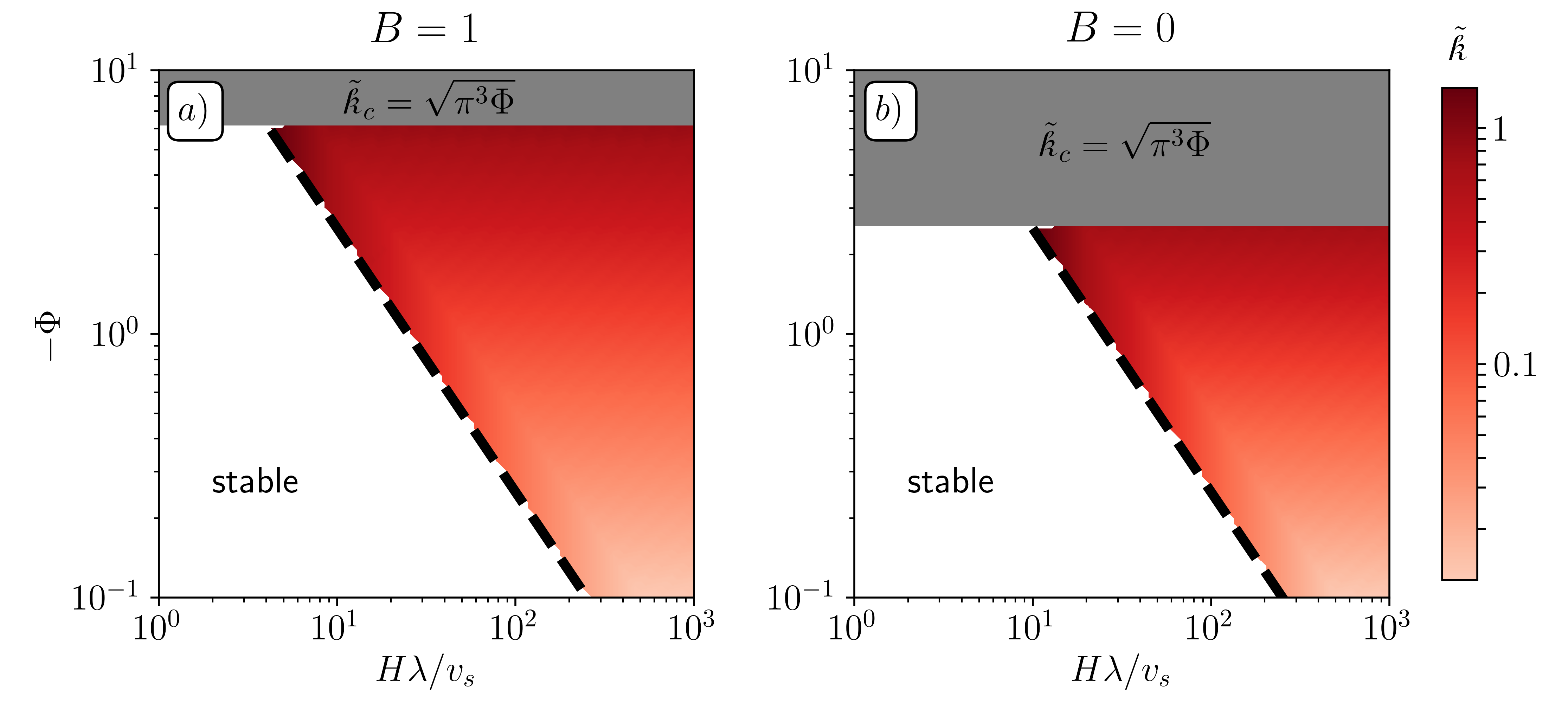}
			\caption{ Density instability in a 2-D puller suspension embedded in a 3-D fluid. The phase diagrams show regions of the two types of density instability obtained by numerical solution of Eq. \eqref{LS2d:eq.denEQ}. At high $\Phi$ (grey regions), the instability sets in at the smallest physically relevant spatial scale, which we assume to be the particle-particle separation, \emph{i.e.}, $\tilde{\kc}=\sqrt{\pi^3 \Phi} \sqrt{v_s^3/\lambda^2 \kappa}$, and set $\sqrt{v_s^3/\lambda^2 \kappa}=1$. At moderate and low $\Phi$ (red heat map), the instability sets in at larger spatial scales set by the maximum of the {\revision arc-like part of} the dispersion law; see Fig. \ref{LS2d:fig.3}. If the system size $H$ is too small, the latter maximum in the dispersion law cannot be accessed, and the instability instead sets in at the scale of the system dimensions. 
               } \label{LS2d:fig.4}
		\end{figure}

    We now proceed by solving Eq. \eqref{LS2d:eq.denEQ} numerically, which we have done for several values of $B$ and $\Phi$, as shown in Fig.~\ref{LS2d:fig.3}. This analysis indeed confirms both  the existence of a {\revision finite}-wavelength instability in puller suspensions at large spatial scales (small $\kc$), and that {\revision a sheet of pushers} is always stable against density perturbations. However, it also shows that, when $\Phi$ is sufficiently negative, $\mathrm{Re}[y]$ increases in an unbounded fashion at large $\kc$ for pullers. Thus, a second density instability, now at small spatial scales, emerges in puller suspensions for sufficiently large densities. To analyse this effect further, we numerically calculate the largest eigenvalue from Eq.~\eqref{LS2d:eq.denEQ}. {The resulting phase diagram is shown in Fig. \ref{LS2d:fig.4}, which displays the critical scale $\kc_c$ as a function of the dimensionless system size $H \lambda/v_s$ and the reduced particle density $\Phi$. Our findings confirm} that the density instability in the confined suspension sets in either at the smallest or the largest available spatial scale, depending on the system parameters as encoded in $\Phi$ and $B$. 
    In drawing the phase diagram we have, in analogy to Section \ref{sec:orientational_instability}, chosen a cutoff at small scales corresponding to the interparticle separation (\emph{i.e.}, $\kc =  \sqrt{ \pi^{3} n }$ ). In the dimensionless units used here, the maximal wave number available is set by $\tilde{\kc}=\sqrt{\pi^3 \Phi} \sqrt{v_s^3/\lambda^2 \kappa}$, where the factor $\sqrt{v_s^3/\lambda^2 \kappa} \sim \mathcal{O}(1)$ for free-swimming \emph{E. coli}  bacteria \citep{Drescher2011}.
    
    {These results show that, while the density instability is present also for spherical particles, the particle shape strongly affects the emergence of small- and large-scale instabilities in that the small-scale instability dominates over a wider range of $\Phi$ values compared to the elongated ($B=1$) particle case. Understanding the physical manifestations of these two instabilities requires detailed investigations using explicit numerical simulations and is left for future work.}
    
	\section{Conclusion}
	\label{Sec:5}
	
	In this work, we have analysed linear stability of the homogeneous and isotropic distribution of microswimmers {\revision restricted} to a two-dimensional plane embedded in a three-dimensional bulk viscous fluid. Our method, based on mean-field kinetic theory, is somewhat different from that used in the previous analysis of bulk suspensions \citep{Saintillan2008, Saintillan2008a,Subramanian2009,Hohenegger2010} and allows us to easily identify the origin of the instability as either an orientational mode, stemming from the mutual reorientation of microswimmers, or a density mode, stemming from local accumulation of microswimmers.
	
	The instabilities we found for {\revision a sheet of microswimmers} are qualitatively different from their bulk counterpart. In such infinite two- or three-dimensional systems, linear stability analysis shows the presence of an orientational instability for pushers above the onset density given respectively by Eqs.~\eqref{LS3d:eq.nc2d} (two-dimensional) and \eqref{LS3d:eq.nc} (three-dimensional). This instability sets in at the largest scale available in the system, typically the dimension of the experimental setup or of the simulation box. Furthermore, due to the incompressible nature of the embedding fluid, dilute two- and three-dimensional bulk suspensions of microswimmers are always stable towards fluctuations in the microswimmer density. 
	
    In contrast, we showed that a two-dimensional layer of microswimmers embedded in a three-dimensional fluid is unstable for both pushers and pullers. For the former swimmer class, an orientational instability similar to the pusher instability in a bulk fluid occurs for large enough densities. However, in contrast to the bulk case, the orientational instability in the two-dimensional layer sets in at the smallest spatial scale, which we interpret as the one below which our continuum description is not valid anymore, \emph{i.e.}, the average particle-particle separation. For parameter values roughly corresponding to \emph{E. coli} bacteria, this leads to the instability setting in at a length-scale of $\sim 5 \mu$m, comparable to the bacterial dimensions.

	On the other hand, a layer of puller microswimmers is susceptible to a density instability characterised by spatial aggregation of microswimmers independent of their orientations. This instability is caused by the mutual advection of the microswimmers and is the direct consequence of the sink-like nature of their in-plane velocity fields, as discussed in Section~\ref{sec:introduction}. In the thermodynamic limit, the puller suspension is unstable at \emph{any} density, while a non-zero critical density emerges for finite system sizes, as shown in Eq.~\eqref{LS2d:eq.denPhic}. The scale at which this puller instability sets in depends on the microswimmer density: at small and moderate densities, the instability occurs at long length scales, while at high enough densities it is replaced by a small-scale instability. While the crossover between these two regimes depends on the particle shape through $B$, both instabilities are generically present regardless of particle shape. Whether this implies that the system reaches different steady states at long times is still an open question that we leave for future work. We also note that the density instability presented here is qualitatively similar to the instability reported by \cite{Baskaran2009}, as both are caused by the fluid's compressibility; in the case of \cite{Baskaran2009}, however, the compressibility arises as a result of an erroneous coarse-graining procedure, as discussed by \cite{Aranson_2022}.
		
	We would like to stress once more that the instability scenarios discussed here are not the consequence of the spatial dimensionality, as two- and three-dimensional bulk suspensions exhibit qualitatively a very similar behaviour. Instead, the qualitative differences to the corresponding bulk suspensions come from the fact that the in-plane fluid in which the microswimmers move is effectively compressible, an effect which is generically present in the case of motion {\revision restricted} to a subset of the embedding space, including microswimmer layers in the vicinity of solid and liquid boundaries. Importantly, this effect is likely to be one of the main driving forces behind the boundary-induced clustering and phase separation observed in self-propelled suspension droplets \citep{Thutupalli2018} and squirmer suspensions \citep{Singh2016}. These authors highlight the importance of the presence of a boundary in inducing effective in-plane attractions among puller swimmers after accumulating at the surface or interface, although they rationalise their findings in terms of attractive hydrodynamic forces without explicitly mentioning the effective compressibility mechanism present in this geometry. Thus, even though our chosen geometry of a two-dimensional plane in an infinite three-dimensional fluid seems artificial, we argue that the generic mechanism demonstrated here is in fact an important driving force behind accumulation in active-particle suspensions near interfaces. 
  	
    Furthermore, the phenomenon of effective in-plane compressibility is not limited to dipolar flow fields. As demonstrated by \cite{Spagnolie2012}, all flow singularities relevant to self-propulsion of microorganisms can be obtained by repeatedly applying the $(\pp \cdot \nabla)$ operator to the Stokeslet and the point source singularity. Therefore, these singularities will have alternating parities with respect to the angle defined by $\xx \cdot \pp$. In the context of the calculation presented in Section~\ref{sec:introduction}, this leads to a layer of force dipoles, force octupoles, \emph{etc.}, having effectively compressible in-plane velocity fields, while the remaining flow singularities remain incompressible even when {\revision restricted} to a lower-dimensional subset of the embedding space. The same reasoning holds for the source singularities, where sources, source quadrupoles, \emph{etc.}, exhibit effectively compressible in-plane velocity fields. Our findings are therefore likely to be of more general importance for hydrodynamic interactions and collective motion of realistic microorganisms near boundaries. 
    
{\revision One of the remaining open questions pertains to the stability of a sheet of microswimmers with respect to out-of-plane perturbations. At first glance, a sheet of pushers generates on average a fluid flow towards the sheet along the third direction and one thus expects out-of-plane perturbations to be suppressed by such a flow. By the same argument, the opposite is expected for pullers. Some support for this hypothesis can be drawn from \cite{Ishikawa2008} who simulated a collection of spherical squirmers confined to a single layer, as in our setup. They observed no instability for pushers, while pullers exhibited formation of dense clusters visually consistent with our density instability. Interestingly, when bottom-heavy puller squirmers were allowed to move out of the original monolayer, the authors observed formation of a dynamical steady state in a form of a microswimmer band, suggesting that pullers are indeed unstable with respect to out-of-plane perturbations. In contrast, the continuum simulations by \cite{Nejad2022} showed coexisting regions of in-plane and out-of-plane director orientations in extensile active nematics, while no such instability was present in the corresponding contractile systems. This complex picture clearly has to be understood \emph{via} a combination of linear stability analysis and fully non-linear simulations, a question that we leave open to future studies. We stress, however, that we view the setup considered in this work as a way to isolate the effect of in-plane compressibility on collective motion of experimental realisations of microswimmers accumulated close to a physical boundary. In such systems, there are always additional effects ensuring microswimmers' presence at the boundary, \emph{e.g.} gravity or surface tension, and the question of stability with respect to out-of-plane perturbations should be addressed in that context.}
    
We conclude by observing that, while the linear stability analysis presented here identifies the origin, the onset conditions, and the associated scales of the instabilities, their non-linear evolution and ensuing non-equilibrium steady states must be assessed in numerical simulations of particle-based models~\citep{Bardfalvy2019,Bardfalvy2020} or mean-field moment equations~\citep{Saintillan2013}. We aim to further explore this and similar questions in future studies.

	
\section*{Acknowledgement}
For the purpose of open access, the authors have applied a Creative Commons Attribution (CC BY) licence to any Author Accepted Manuscript version arising from this submission.

\section*{Funding} 
JS acknowledges financial support from the Swedish Research Council (Project No. 2019-03718). AM acknowledges financial support from EPSRC (grant number EP/V048198/1). CN acknowledges funding from Institut de physique du CNRS.

\section*{Competing Interestst}
The authors report no competing interests.

	\appendix
	

	\section{Dipolar fluid velocity} 
	\label{App:ud}
	
	The far-field fluid velocity $\uu$ generated by a swimming microorganism at low Reynolds number is described by the hydrodynamic force dipole,  \citep{Lauga2009,Spagnolie2012}. The latter satisfies Stokes' equation driven by two point-like forces $\pm f \pp$ acting on the fluid at the positions $ \pm \tfrac{1}{2}l\pp $,
	\begin{align}
	&\mu \nabla^{2} u^{\alpha}(\xx) - \nabla^{\alpha} P(\xx) + f p^{\alpha} [ \delta(\xx- \tfrac{1}{2} l \pp) - \delta(\xx+ \tfrac{1}{2} l \pp)] = 0, 
	\label{App:ud.1}\\
	&\qquad\qquad\qquad\qquad \nabla^{\alpha} u^{\alpha}(\xx) = 0, 
	\label{App:ud.2}
	\end{align}
	Here, $\mu$ is the viscosity of the fluid, $P$ is the pressure, and $\delta(\xx)$ denotes the Dirac delta-function. The solution to this system is readily found by performing the $d$-dimensional Fourier transform,  \eqref{KT:eq.FT}, which gives
	\begin{align}
	& -\mu k^{2} \hat{u}^{\alpha} - i k^{\alpha} \hat{P} + f p^{\alpha}[\ee^{-\frac{1}{2}i l \kk \cdot \pp} - \ee^{\frac{1}{2}i l \kk \cdot \pp}] = 0, 
	\label{App:ud.3} \\
	& \qquad\qquad\qquad\qquad  k^{\alpha} \hat{u}^{\alpha} = 0. 
	\label{App:ud.4}
	\end{align}
	The point-dipole approximation, relevant at large scales where $ k l \ll 1  $, is obtained to linear order in the dipolar length $l$, yielding
	\begin{align}
	\hat{u}^{\alpha}(\kk,\pp) = - i \kappa \frac{\kk\cdot\pp}{k^{2}} \PP^{\alpha\beta} p^{\beta}.
	\label{App:ud.ud3K}
	\end{align}
	Here, $ \PP^{\alpha\beta} = \delta^{\alpha\beta} - k^{\alpha}k^{\beta}/k^{2} $ is the transversal projection operator, $k=\vert \kk \vert$, and  $ \kappa = f l / \mu $ is the dipolar strength.
	
	The derivation presented above is independent of the spatial dimensionality $d$ and the Fourier transform of the dipolar field \eqref{App:ud.ud3K} has the same form for bulk systems of all dimensions. Its real space representation, however, strongly depends on $d$. In three dimensions, the inverse Fourier transform \eqref{KT:eq.FT} yields
	\begin{align}
	u_{3d}^{\alpha}(\rr,\pp) = \frac{\kappa}{8\pi} \frac{r^{\alpha}}{|\rr|^{3}} \left[ 3 \frac{(\rr\cdot\pp)^{2}}{|\rr|^{2}} - 1 \right], \quad \rr = \xx-\xx_{0}, 
	\label{App:ud.ud3X}
	\end{align}
	where we have introduced an arbitrary position $ \xx_{0} $ of the force dipole. In two dimensions, the real-space representation is
	\begin{align}
	u_{2d}^{\alpha}(\rr,\pp) = \frac{\kappa}{4\pi} \frac{r^{\alpha}}{|\rr|^{2}} \left[ 2 \frac{(\rr\cdot\pp)^{2}}{|\rr|^{2}} - 1 \right], \quad \rr = \xx-\xx_{0}, 
	\label{App:ud.ud22X}
	\end{align}
	where we notice that the dipolar strength $\kappa = f l / \mu$ has different physical dimensions in two- and three-dimensional bulk systems.
	
	The velocity field of a hydrodynamic dipole {\revision restricted} to a two-dimensional plane embedded in an infinite three-dimensional fluid can be obtained from \eqref{App:ud.ud3X}. 
    We select $z=0$ to be the plane of the microswimmers, and recall the definitions of two-dimensional vectors $\xcb=(x,y)$, and $\pcb=(p_x,p_y)$, from the main text. Performing the corresponding two-dimensional Fourier transform of  \eqref{App:ud.ud3X} with respect to $\xcb$, leads to
	\begin{align}
	\hat{u}_{\text{plane}}^{\alpha}(\kcb,\pcb) = - \frac{i\kappa}{2} \frac{\kcb\cdot\pp}{\kc} \left[ \PP^{\alpha\beta} + \frac{1}{2} \QQ^{\alpha\beta} \right] p^{\beta}, 
	\label{LS2d:eq.u2dKb}
	\end{align}
	where $ \kcb = \left( \kc_x, \kc_y \right)$ is the wavevector associated to $\xcb$, and we have introduced the longitudinal projection operator $\QQ^{\alpha\beta} = \kc^{\alpha}\kc^{\beta}/\kc^{2}$.
	
	\section{Linear stability analysis of 2-D bulk suspensions}
	\label{App:2D}
	
	Here, we repeat the calculation presented in Section \ref{Sec:3} for the case of microswimmers suspended in two-dimensional bulk fluid. We set $d = 2$ in \eqref{LS:eq.LinEq1}--\eqref{LS:eq.LinEq2}, and employ \eqref{App:ud.ud3K}, since the form of the dipolar Fourier transform is the same in all dimensions, as noted in Appendix \ref{App:ud}. We obtain the following eigenvalue problems
	\begin{align}
	\delta \hat{\rho} &= \frac{\lambda}{v_{s} k} \frac{b}{\sqrt{1+b^{2}}} \delta \hat{\rho}, 
	\label{LS3d:eq.rho2d} \\
	\delta \hat{\UU}^{\alpha} &=\frac{B n \kappa}{v_{s}k} \frac{b^{2}-  2 \sqrt{1+b^{2}} + 2}{b^{3}} \delta \hat{\UU}^{\alpha}.
	\label{LS3d:eq.U2d}
	\end{align}
    which, similarly to the three-dimensional case, are decoupled.
	The orientational eigenvalue problem can be re-written as 
	\begin{align}
	\gamma_{2d} \equiv \frac{v_{s}k}{B n \kappa} = \frac{b^{2}-  2 \sqrt{1+b^{2}} + 2}{b^{3}}, 
	\label{LS3d:eq.aG}
	\end{align}
	where, as before, $b = v_{s}k/(\chi + \lambda)$. In contrast to \eqref{LS3d:eq.aF}, the solution to \eqref{LS3d:eq.aG} can be found analytically, yielding
	\begin{figure}
		\centering
		\includegraphics[width=0.9\linewidth]{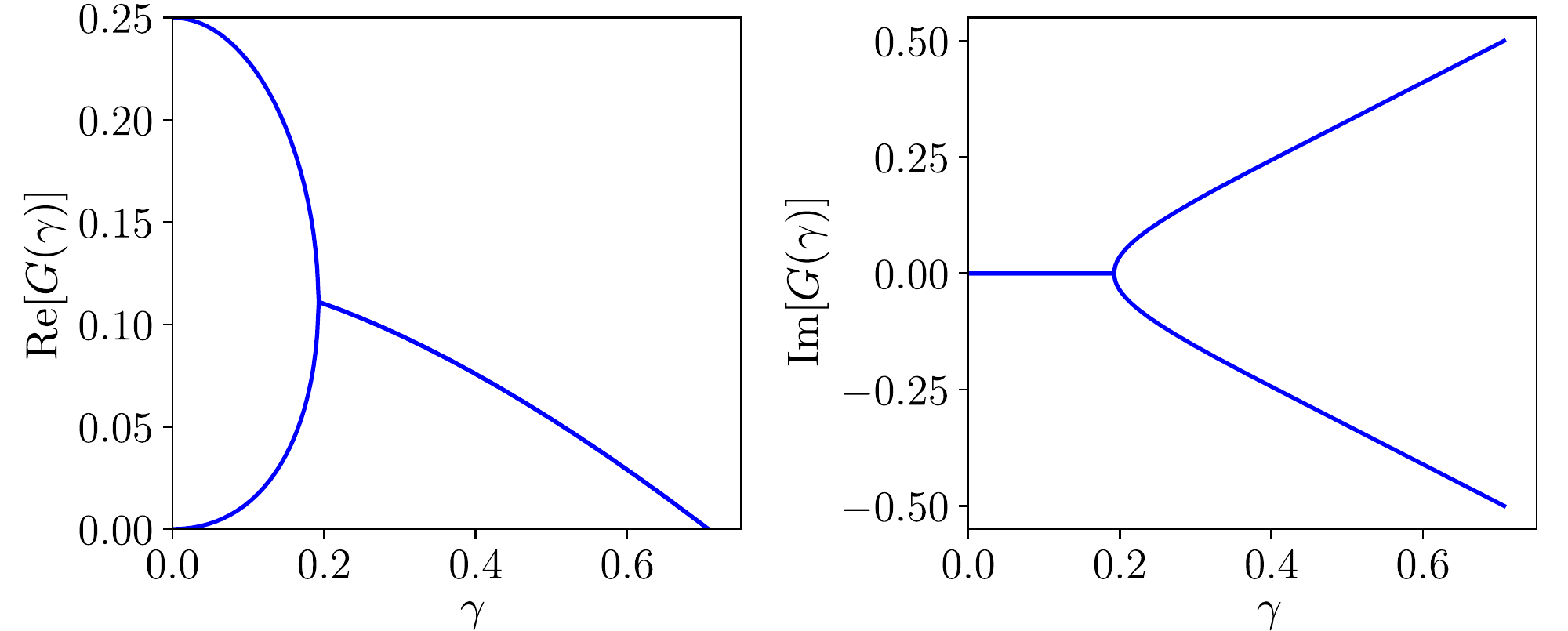}
		\caption{The real (left) and imaginary (right) part of the function $ G(\gamma) $ from \eqref{LS2d:eq.G}. } 
		\label{LS2d:fig.G}
	\end{figure}
	\begin{align}
	\chi = - \lambda + G(\gamma_{2d}) B \kappa n,
	\label{LS3d:eq.aGG}
	\end{align}
where $G(\gamma_{2d})$ is given in Eq. \eqref{LS2d:eq.G}.
The plot of $ G$ is shown in Fig. \ref{LS2d:fig.G}, and it has the same qualitative features as $F(\gamma)$ shown in Fig. \ref{LS3d:fig.1}. As in the 3-D case, the orientational instability only exists for pushers, $\kappa>0$, and sets in at the largest possible scale, $ k \rightarrow 0 $. The corresponding instability condition is given by
	\begin{align}
	n > n_c = \frac{4\lambda}{B\kappa}, \label{LS3d:eq.nc2d}
	\end{align}
	where the number density $n$ and the dipolar strength $\kappa$ have different dimensions than their 3-D counterparts. The solution \eqref{LS3d:eq.aGG} ceases to exist for $\gamma_{2d}>1/\sqrt{2}$, where one can instead prove stability using an analysis completely analogous to the 3-D case (see Appendix~\ref{App:2D.LT}). 
The analysis of density fluctuations is significantly simpler and the corresponding eigenvalue is readily obtained from \eqref{LS3d:eq.rho2d}:
\begin{align}
	\chi = - \lambda \pm \sqrt{\lambda^{2} - k^{2} v_{s}^{2}}.
	\end{align}
	Clearly, $\chi$ is negative for all $k$, and thus the suspension is always stable with respect to small fluctuations in microswimmer density. 
	
	\section{Linear stability analysis using Laplace transforms}

    \subsection{Density fluctuations in bulk suspensions}\label{App:2D.LT}
	
     As mentioned in Section \ref{sec:orientational_instability}, the linear stability analysis based on the Ansatz \eqref{LS:eq.SELinExp} leads to a missing solution of the eigenvalue problem at small spatial scales. This issue can be tackled by exactly solving the linear dynamics  \eqref{LS:eq.SELin} via Laplace transforms. We introduce the following notation for the forward Laplace transform
	\begin{align}
	f(s) &= \int_{0}^{\infty} \dd t f(t) \ee^{-st} \label{LT:eq.LT}
	\end{align}
    where $s$ has a positive real part. For the inverse Laplace transformation, we use the definition
	\begin{align}
	f(t) = \frac{1}{2\pi i} \lim\limits_{T\rightarrow\infty} \int_{\gamma-i T}^{\gamma+i T} \dd s \ f(s) \ee^{st}, \label{inv}
	\end{align}
	where $ \gamma $ must be chosen such that the real part of every pole in $ f(s) $ is smaller than $ \gamma $.
 
    We now consider the case of a 3-D bulk suspension, for which the  Laplace transform of \eqref{LS:eq.SELin} gives
	\begin{align}
	(s  + \lambda + i v_{s} \pp\cdot\kk) \delta \hat{\Psi} -  \delta \hat{\Psi}_{0} &= \frac{\lambda}{4\pi}  \delta \hat{\rho}  +  \frac{3Bn}{4\pi} ik^{\alpha}  p^{\alpha} p^{\beta} \delta \hat{\UU}^{\beta}, 
	\end{align}
	where $\delta \hat{\Psi}_{0} = \delta \hat{\Psi}(\kk,\pp,t=0) $ represents the initial condition. Dividing the last equation with the linear operator on the left hand side and integrating over the orientation $\pp$ leads to
	\begin{align}
	\delta \hat{\rho}(s) &=  \frac{1}{1 - \frac{\lambda}{v_{s}k}\arctan(\frac{v_{s}k}{s + \lambda})} \int \dd \pp \ \frac{\delta \hat{\Psi}_{0}}{s  + \lambda + i v_{s} \pp\cdot\kk}. \label{App:2D.LT_rho} 
	\end{align}
     The expression \eqref{App:2D.LT_rho} contains two separate poles
	\begin{align}
	s_{1} &= - \lambda - i v_{s} \pp\cdot\kk, \quad  s_{2} = - \lambda + \frac{v_{s}k}{\tan(\frac{v_{s}k}{\lambda})}, 
	\end{align}
	where the second pole appears only in the limited region of parameter space $ v_{s}k/\lambda < \pi/2 $. The result of the inverse Laplace transformation \eqref{inv} of \eqref{App:2D.LT_rho} consists of two terms, one corresponding to each pole:
	\begin{align}
	\delta \hat{\rho}(t) = \delta \hat{\rho}_{1}(t) + \delta \hat{\rho}_{2}(t)
 \label{App:LT.eq.rho}
	\end{align}
    which we now analyse separately. 
	The first term in \eqref{App:LT.eq.rho} exists for any $ v_{s}k/\lambda $ and reads
	\begin{align}
	\delta \hat{\rho}_{1}(t) &= 
	\int \dd \pp \ \Bigg [ \frac{1}{1 - \frac{\lambda}{v_{s}k}\arctan(\frac{i k}{\pp\cdot\kk})} \ee^{- \left( \lambda + iv_{s}\pp\cdot\kk \right)t} \Bigg] \delta \hat{\Psi}_{0}(\kk,\pp). \label{App:LT.eq.rho1}
	\end{align}
The integral over $\pp$ is conveniently expressed in spherical coordinates with its polar axis being along $\kk$, and reads
\begin{align}
\delta \hat{\rho}_{1}(t) &= \ee^{-\lambda t} \int_{-1}^{1} \dd x \ F(x) \ee^{- iv_{s}k t x }, \quad F(x) =  \frac{\int_{0}^{2\pi} \dd \phi \ \delta \hat{\Psi}_{0}(\kk,\phi,\arccos(x))}{1 - \frac{\lambda}{v_{s}k}\arctan(\frac{i}{x})}. 
\label{App:LT.eq.rho1b}
\end{align}
We observe that $\int_{-1}^{1} \dd x | F(x) |<\infty$, provided $\max_x \Big|\int_{0}^{2\pi} \dd \phi \ \delta \hat{\Psi}_{0}(\kk,\phi,\arccos(x))\Big| < \infty$, as expected from a physical initial condition $\delta \hat{\Psi}(\kk,\pp)$. Applying the Riemann-Lebesgue lemma \citep{BenderOrszag}, we conclude that $\delta \hat{\rho}_{1}(t) \to 0$, as $t\to\infty$ for $v_s k>0$.
	

	The second term in \eqref{App:LT.eq.rho} exists only for $ v_{s}k/\lambda < \pi/2 $ and has the form
	\begin{align}
	\delta \hat{\rho}_{2}(t) &=  \exp\left[- \left( \lambda-\frac{v_{s}k}{\tan(v_{s}k/\lambda)} \right)t \right]
	\int \dd \pp \ \Bigg[ \frac{(\frac{v_{s}k}{\lambda})}{\tan(\frac{v_{s}k}{\lambda})} \frac{1 + \tan^{2}(\frac{v_{s}k}{\lambda})}{1 + i\tan(\frac{v_{s}k}{\lambda})\pp\cdot\kk }   \Bigg] \delta \hat{\Psi}_{0}(\kk,\pp). \label{App:LT.eq.rho2}
	\end{align}
	The temporal evolution of \eqref{App:LT.eq.rho2} follows an exponential decay provided that
	\begin{align}
	\lambda - \frac{v_{s}k}{\tan(v_{s}k/\lambda)} > 0,
	\end{align}
	which always holds for $ v_{s}k/\lambda < \pi/2$. $\delta\hat{\rho}_2$ thus corresponds to the solution obtained when using the exponential Ansatz $ \delta\hat{\rho}(t) \sim \exp\{ s t \} $, as was analysed in Section \ref{LS2D:eq.DebInst}.

 We conclude that, although the use of the exponential Ansatz in \eqref{LS:eq.SELinExp} does not allow for retrieving the sub-exponential contributions to the linear dynamics, the missing part does not carry any new information relevant for the linear stability analysis.

\subsection{Orientational instability in {\revision a sheet of microswimmers}}\label{App:2D.LT_2d}

We now repeat the calculation for the case of transversal velocity fluctuations in a two-dimensional suspension {\revision restricted} to a plane in a three-dimensional fluid. The starting point is the Smoluchowski equation \eqref{LS:eq.SELin} in 2-D:
	\begin{align}
	\Bigl[ \partial_{t}  + \lambda + i v_{s} \pcb\cdot\kcb \Bigr] \delta \hat{\Psi} &= \frac{\lambda}{2\pi}  \delta \hat{\rho}  +  \frac{ n}{2\pi} \Bigl[ 2 B \pc^{\alpha} \pc^{\beta}  - (1+B) \delta^{\alpha\beta}  \Bigr] i \kc^{\alpha} \delta \hat{\UU}^{\beta},  \label{LT:eq.SE2d}
	\end{align}
 Eq. \eqref{LT:eq.SE2d} is now transformed using \eqref{LT:eq.LT}, and after multiplying with $\PP^{\alpha\beta}\hat{u}^{\beta}_{\text{plane}}$ and integrating over the orentation $\pcb$, we arrive at
\begin{align}
    \delta \hat{\UU}_{\perp}^{\alpha}(s) = \frac{1}{1 - M_{11}} \int_{0}^{2\pi} \frac{\dd \theta}{2\pi} \frac{ \PP^{\alpha\beta} \hat{u}_{\text{plane}}^{\beta} \delta \hat{\Psi}_0(\kcb,\theta)}{s + \lambda + i v_s \kc \cos\theta}, \label{LT:eq.Ut}
\end{align}
where $M_{11}$ is defined in \joakim{\eqref{LS:eq.M2}}. Similarly to before, the inverse Laplace transform consists of two parts, each corresponding to one of the poles in Eq. \eqref{LT:eq.Ut}. The first one is the pole corresponding to $M_{11}=1$, which leads to the dispersion law in Eq.~\eqref{LS2d:eq.yd}. As discussed in the main text, the latter solution ceases to exist for $ 2\sqrt{2} v_{s} > B \kappa n $. In this region, the stability is instead determined by the second pole $s = -\lambda - i v_{s} \kc \cos \theta$, where the time representation of the corresponding solution is given by
\begin{align}
    \delta \hat{\UU}_{\perp}^{\alpha}(t) = \ee^{-\lambda t}
	\int_{0}^{2\pi} \dd \theta \ F(\theta) \ee^{- iv_{s} \kc t \cos\theta }, \quad F(\theta) = \frac{\PP^{\alpha\beta} \hat{u}_{\text{plane}}^{\beta} \delta \hat{\Psi}_0(\kcb,\theta)}{1 - M_{11}|_{b\rightarrow\frac{i}{\cos\theta}}}. 
\end{align}
Again, it is possible to show that $\int_{0}^{2\pi} \dd \theta | F(\theta) |$ is bounded provided $2\sqrt{2} v_s > B\kappa n$ and $ \delta \hat{\Psi}_0(\kcb,\theta)$ satisfies 
the same physical requirements as in Appendix \ref{App:2D.LT}. 
Applying the Riemann-Lebesgue lemma, we conclude that  $\delta \hat{\UU}_{\perp}^{\alpha}(t)$ vanishes as $t \to \infty$, and the system is stable for $2\sqrt{2} v_s > B\kappa n$.
 
	\section{Approximation of $n_c$ for {\revision a sheet of pushers}} \label{App:2d_approx}
	
\begin{figure}
		\centering
		\includegraphics[width=0.6\linewidth]{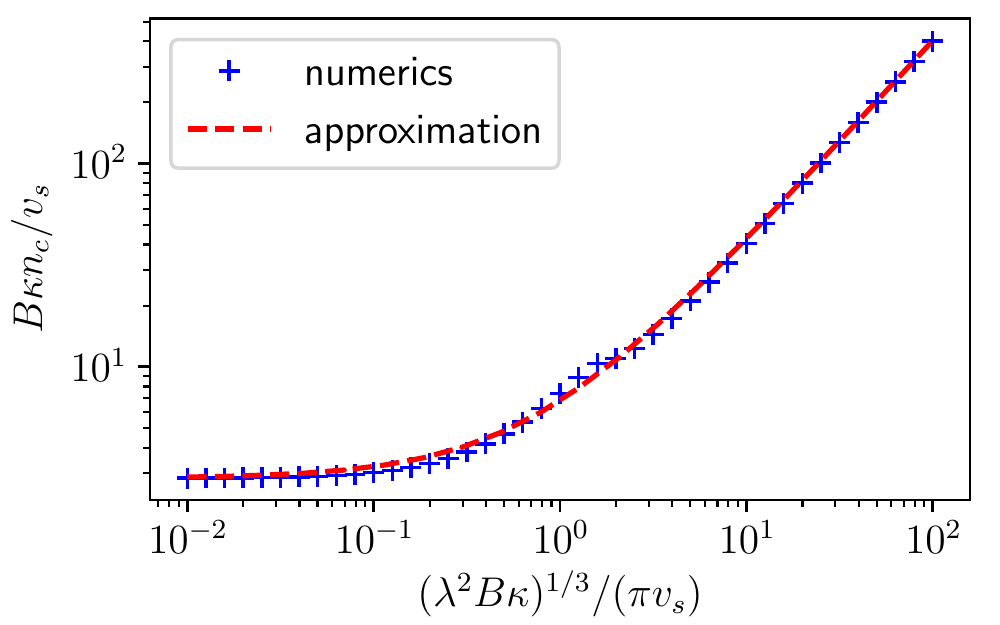}
		\caption{Comparison of the approximation \eqref{LS2d:eq.OrientPhic1} with the exact values obtained by numerical solution of $ \text{Re}[\chi] = 0 $, where $\chi$ is given by Eq. \eqref{App:2d_approx.y}. Our approximation shows good quantitative agreement with the exact numerical values. The kink in the numerical data is not an artefact and corresponds to a switch between two eigenvalue branches.} \label{App:2d_approx:fig}
	\end{figure}

	In this Appendix, we will derive the approximate expression  \eqref{LS2d:eq.OrientPhic1} for the critical density $n_c$ corresponding to the orientational instability setting in at the scale of particle-particle separation in a 2-D layer of pusher microswimmers. We set $\kc_{c} = \sqrt{\pi^{3}n_{c}}$ and $n=n_c$ in \eqref{LS2d:eq.yd}, giving
	\begin{align}
	\chi = - \lambda + G\left(\frac{2 v_{s}}{B\kappa n_{c}}\right) \frac{B \kappa (\pi n_{c})^{3/2}}{2}. \label{App:2d_approx.y}
	\end{align}
	We now separate the expression into two approximations valid asymptotically for small and large values of $v_{s}$. We first set $ v_{s} \rightarrow 0 $ in \eqref{App:2d_approx.y}, yielding
	\begin{align}\label{eq:cn-2}
	\chi(v_s \to 0) = - \lambda + \frac{1}{8} B \kappa (\pi n_{c})^{3/2},
	\end{align}
	where $ \text{Re}[\chi] = 0 $ now gives the `shaker' approximation
    \begin{align}
    n_{c}(v_s \to 0) = \frac{4}{\pi} \left( \frac{\lambda}{B \kappa} \right)^{2/3}. \label{App:eq.vs0}
    \end{align}
    For the fast-swimming limit, we use the result \eqref{LS2d:eq.yd2}, \emph{i.e.}, we assume that the critical density is linearly proportional to the swimming speed, 
    \begin{align}
    n_{c}(v_s \to \infty) = 2\sqrt{2} \frac{v_{s}}{B\kappa} + \mathcal{O}(1). \label{App:eq.vsInf}
    \end{align}
	Combining \eqref{App:eq.vs0} and \eqref{App:eq.vsInf} gives the result \eqref{LS2d:eq.OrientPhic1}. The latter shows a good quantitative agreement with the numerical solution of $ \text{Re}[\chi] = 0 $, as shown in Fig. \ref{App:2d_approx:fig}.

	\bibliographystyle{jfm}
	\bibliography{refs}
	
\end{document}